\PassOptionsToPackage{dvipsnames}{xcolor}
\documentclass[longbibliography,nofootinbib,aps,prx,superscriptaddress,twocolumn]{revtex4-2}

\usepackage{amsmath,amsfonts,amssymb,amsthm,dcolumn,bm,qcircuit,appendix,mathtools,thmtools,thm-restate,mathrsfs,pgfplots,soul,lipsum,graphicx,braket,enumitem,caption,subcaption,ragged2e}

\usepackage{tikz,tkz-euclide,tikz-3dplot}
\usetikzlibrary{positioning}

\usetikzlibrary{matrix,fit,backgrounds,3d,arrows, decorations.pathreplacing,shapes.geometric,3d, calc}
\usetikzlibrary{arrows.meta,calc}
\tikzset{
yellowv/.style={circle,draw,fill=yellow!80!orange,inner sep=2.2pt},
whitev/.style={circle,draw,fill=white,inner sep=1.9pt},
e/.style={line width=0.8pt},
}
\def\BibTeX{{\rm B\kern-.05em{\sc i\kern-.025em b}\kern-.08em
    T\kern-.1667em\lower.7ex\hbox{E}\kern-.125emX}}
    
\usepackage[ruled,vlined]{algorithm2e}
\usepackage{algorithmic}

\newtheorem{theorem}{Theorem}
\newtheorem{lemma}{Lemma}

\newtheorem{remark}{Remark}
\newtheorem{definition}{Definition}

\newcommand{\xbf}{\textbf{x}}

\newcommand{\Ibb}{\mathbb{I}}

\newcommand{\Rbb}{\mathbb{R}}

\newcommand{\Zbb}{\mathbb{Z}}
\newcommand{\Scal}{\mathcal{S}}
\newcommand{\norm}[1]{\left\lVert#1\right\rVert}

\usepackage{xcolor}
\usepackage{hyperref}

\hypersetup{
	colorlinks = true,
	linkcolor = [rgb]{0.70,0.13,0.13},
	citecolor = [rgb]{0.13,0.55,0.13},
	urlcolor = [rgb]{0.25, 0.41, 0.88}}
\usepackage[capitalize,nameinlink]{cleveref}

\begin{document}
\title{Quantum topological data analysis algorithm for dynamical systems}

\author{Nhat A. Nghiem}
\email{{nhatanh.nghiemvu@stonybrook.edu}}
\affiliation{Department of Physics and Astronomy, State University of New York at Stony Brook, Stony Brook, NY 11794-3800, USA}
\affiliation{C. N. Yang Institute for Theoretical Physics, State University of New York at Stony Brook, Stony Brook, NY 11794-3840, USA}

\begin{abstract}
Dynamical systems appear in nearly every aspect of the physical world. As such, understanding the properties of dynamical systems is of great importance. Typically, a dynamical system is described by a system of ordinary differential equations (ODE). Most ODEs do not admit analytical solutions, which makes dynamical systems challenging to understand. In this work, we introduce a quantum framework for determining certain properties of dynamical systems. We combine many recent advances in quantum algorithms, particularly quantum ODE solver and quantum topological data analysis. Leveraging the prior results regarding the quantum ODE solvers, we use the output of these quantum algorithms as a means to build the graph associated with the trajectory of the dynamical system in the phase space. This graph information is then fed into existing quantum TDA algorithms, respectively, to obtain the (normalized) Betti numbers, which are the topological signatures. We then discuss how such signatures can be linked to the properties of a given ODE, and thus, they can reveal insight towards the original dynamical systems. As a by-product, our work provides an affirmative answer to the applicability of quantum ODE solvers, showing that the quantum state as output can be valuable for useful purposes. 

\end{abstract}

\maketitle

\section{Introduction}
Quantum computation has emerged as a new type of technology that can potentially reshape the computational frontier. By harnessing the intrinsically quantum features, such as entanglement and superposition, quantum computers hold the promise to deliver solutions to challenging computational problems that may lie beyond the reach of classical devices. As such, the development of quantum algorithms for tackling problems has seen a major progress. For example, early attempts have shown that quantum computers can efficiently factorize a large number \cite{shor1999polynomial}, search an unstructured database \cite{grover1996fast}, computing black-box function properties \cite{deutsch1985quantum,deutsch1992rapid}. Simulating quantum systems has also been regarded as the most natural scenarios for achieving quantum speedups \cite{berry2007efficient,berry2012black,berry2014high,berry2015hamiltonian,berry2015simulating, childs2022quantum,bauer2023quantum, childs2018toward, lloyd1996universal, gerritsma2010quantum}. In recent years, quantum algorithms for machine learning and data science problems have received significant attention \cite{harrow2009quantum,childs2017quantum, lloyd2013quantum,lloyd2016quantum, lloyd2020quantum, mitarai2018quantum, schuld2018supervised,schuld2019evaluating,schuld2019machine,schuld2019quantum,schuld2020circuit, schuld2020effect}. In a similar manner, quantum algorithms for partial differential equations and ordinary differential equations have gathered many efforts  \cite{leyton2008quantum, liu2021efficient, zanger2021quantum, berry2014high, bagherimehrab2023fast, krovi2023improved, liu2023efficient, childs2020quantum, jin2024quantum, childs2021high, linden2022quantum, tanaka2023polynomial, shang2024design, tennie2025quantum, jin2025schrodingerization}. Most recently, there is a surge of interest in applying quantum computation to topological data analysis (TDA) \cite{lloyd2016quantum, schmidhuber2022complexity, berry2024analyzing, hayakawa2022quantum, hayakawa2024quantum, nghiem2023quantum, nghiem2025hybrid, lee2025new}. 

In this work, we continue the above line of pursuit. Specifically, we are curious about how quantum computers can be useful in probing dynamical systems. Dynamical system is a major object of study, especially in mathematics and physics. However, it is arguably fair to say that dynamical systems appear in almost every area, ranging from natural science to social science. A pendulum, or a system of mass-spring forms the simplest dynamical systems. A system of many mass-spring coupled to each other form a more complex dynamical system. Dynamical systems typically involve ordinary differential equations (ODE), and thus the ability to solve ODEs is often critical to understanding the features of dynamical systems. For example, for a simple mass-spring setup, it is well-known that the mass will oscillate, as in this case its coordinates can be solved exactly. However, for many mass-springs system, it is not as straightforward, even if we limit to just 1-dimensional. Higher dimension is, therefore, possess a major challenge to understand the dynamical properties. 

Motivated by the previous attempts on quantum algorithms, we combine the insights from quantum ODE solvers and quantum TDA algorithms. We particularly show that the output of quantum ODE solvers, which are typically a quantum state that contains the solution to some given ODE, can be used to construct the input to quantum TDA algorithms. The output of quantum TDA algorithms are then shown to be connected to the dynamical properties of the original dynamical system. This result in turn provides another positive example toward an important question in the field, which concerns whether the quantum state representation of the output of quantum algorithms can be useful. This question has somewhat deferred the exponential speedup claimed by previous quantum algorithm efforts, casting certain doubt on the potential application of quantum computers in real-world problems 

Our work is organized as follows. In Section \ref{sec: interplay}, we discuss in detail the interplay between dynamical system and topological data analysis. We divide such section into three subsections. The first one, Sec.~\ref{sec: TDA} covers a brief introduction to TDA. The second one, Sec.~\ref{sec: overviewdynamicalsystem} describe a general view on dynamical system. Section \ref{sec: dynamicalviaTDA} shows how dynamical system can be viewed from TDA perspective. We then proceed to outline our main proposal in Section \ref{sec: qTDAfordynamicalsystem}. Conclusion is then given in Section \ref{sec: conclusion}.

\section{Interplay between Dynamical system and topological data analysis}
\label{sec: interplay}
In this section, we provide a few preliminaries of our work, including some backgrounds on topological data analysis (TDA), dynamical system, and eventually the interplay between them.

\subsection{Topological data analysis (TDA)}
\label{sec: TDA}
\begin{figure}
    \centering
    \begin{tikzpicture}[scale=2, line join=round]
\coordinate (P0) at (0,1);
\coordinate (P1) at (-0.8,0);
\coordinate (P2) at (0.4,-0.2);
\coordinate (P3) at (0.4,0.4);
\filldraw[black] (P0) circle (1pt);
\filldraw[black] (P1) circle (1pt);
\filldraw[black] (P2) circle (1pt);
\filldraw[black] (P3) circle (1pt);

\draw[thick] (P0) -- (P1) -- (P2) -- (P0);
\draw[fill = gray!30,thick] (P0) -- (P1) -- (P2);
\draw[thick,fill=gray!10] (P0) -- (P3) -- (P2);
\draw[dashed, thick] (P1) -- (P3);
\draw[thick] (P0) -- (P2);
\node[above] at (P0) {$p_0$};
\node[left] at (P1) {$p_1$};
\node[below right] at (P2) {$p_2$};
\node[right] at (P3) {$p_3$};
\end{tikzpicture}
    \caption{0-simplex is a point, 1-simplex is a line, 2-simplex is a triangle and a 3-simplex is illustrated by the figure above. }
    \label{fig: simplex}
\end{figure}
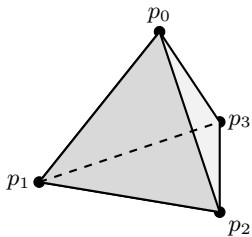

\noindent
\textbf{Algebraic topology.} We begin with a remark that a more self-contained overview of algebraic topology can be found in the Appendix \ref{sec: reviewofalgebraictopology}. Here, we recapitulate a few essential concepts and notations for our subsequent construction. Throughout our work, we use $\sigma_{r_i}$ to denote the $i$-th $r$-simplex and $\Scal_r$ to denote the set of $r$-simplexes. An illustration for 3-simplex is given in Figure \ref{fig: simplex}, with the generalization to higher dimension being straightforward. A $r$-chain $c_r$ is a linear combination of $r$-simplexes, which is of the form $\sum_i a_i \sigma_{r_i}$. In principle, $\{a_i\}$ can be from real field $\Rbb$, or integer $\Zbb$. In our work, we choose real field, for simplicity. In such a case, the set of $r$-chains $\{c_r\}$ form a namely $r$-chain vector space $C_r$ with $r$-simplexes $\{ \sigma_{r_i}\} $ as basis. The $r$-boundary operator $\partial_r$ is a linear operator that acts on $r$-simplex $[v_0, v_1, \ldots, v_r]$ as follows: 
\begin{equation}
\partial_r [v_0, v_1, \ldots, v_r] = \sum_{i=0}^r (-1)^i [v_0, v_1, \ldots, \hat{v_i}, \ldots, v_r],
\end{equation}
where $\hat{p_i}$ indicates that vertex $p_i$ is omitted, yielding an $(r{-}1)$-simplex. The action of $\partial_r$ on $r$-chain $c_r = \sum_i a_i \sigma_{r_i}$ is extended via linearity, e.g., $\partial_r ( \sum_i a_i \sigma_{r_i}) =  \sum_i a_i \partial_r\sigma_{r_i}$. The boundary operators form a chain complex:
\begin{equation}
0 \xrightarrow{} C_n \xrightarrow{\partial_n} C_{n-1} \xrightarrow{\partial_{n-1}} \cdots \xrightarrow{\partial_1} C_0 \xrightarrow{\partial_0} 0.
\end{equation}
A $r$-chain $\sigma_r$ is called a cycle if $\partial_r \sigma_r = 0$. At the same time, it is called a $r$-boundary if there is a $(r+1)$-chain $c_{r+1}$ such that $\partial_{r+1} c_{r+1} = \partial_r$. A simplicial complex $\mathcal{K}$ is a collection of simplexes such that the intersection between any pair of simplexes is either empty or another simplex $\in \mathcal{K}$. 

An important concept within algebraic topology is the Betti number. The $r$-th Betti number $\beta_r$ is defined as the dimension of the $r$-th homology space $H_r \equiv \rm Ker \  \partial_r/ \rm Im \  \partial_{r+1}$. This number essentially counts the number of $r$-dimensional ``holes'' in $\mathcal{K}$. In particular, the set of Betti numbers $\beta_1,\beta_2,...$, etc, is the topological invariant associated with the given complex $\mathcal{K}$. If two different simplicial complexes have the same set of Betti numbers, then they are regarded as the triangulations of the same topological space. As a remark, this is how algebra intertwines with topology, where the topological structure is reflected via algebraic structure. \\

\noindent
\textbf{Topological data analysis (TDA).} TDA is an emerging area \cite{wasserman2016topological} that applies algebraic topology to analyze large-scale data, which are usually provided as data points, or point cloud, possibly with edges connecting them. This type of setting emerges naturally within many areas, e.g., network analysis, medical imaging. By treating a vertex as a 0-simplex, an edge as a 1-simplex, and a triangle (3 vertices with 3 edges connecting them) as a 2-simplex, and so on, a simplicial complex can be identified with the given dataset. As a result, one can leverage key results and related linear algebraic techniques to probe the structure of such complex, which in turn, provide the insight into the original point cloud. For example, by estimating the Betti numbers, which captures the number of ``holes'', one can obtain insight about the shape of the given dataset. As a comment, algebraic topology inherently incorporates high-dimensional objects (e.g. 2-simplex and above), which offers a more versatile and thorough analytical tool, compared to the traditional graph-theoretic techniques that is only concerned with vertex and edge.

\subsection{Dynamical system}
\label{sec: overviewdynamicalsystem}
A dynamical system describes how a state $\xbf$ evolves over time according to fixed rules. It is defined by two essential concepts. \textbf{State space:} all possible configurations, usually Euclidean space $\Rbb^n$ or a manifold. \textbf{Evolution law:} A rule that tells you the next state. With respect to the evolution law, there are typically two manners. The continuous-time evolution is characterized by a system of ordinary differential equations (ODE), or partial differential equation (PDE) $  \frac{d \xbf}{dt} = F(\xbf,t)$ where $\xbf \equiv \xbf(t)$ changes smoothly with time. At the same time, discrete-time evolution is featured with a sequence of discrete values $   \xbf_{t+1} = F(\xbf_t)$ where $F(.)$ generally being some function.

Dynamical systems are, in fact, everywhere. A spring-mass system, a pendulum composed of a string and mass, a physical system with interacting components, etc, are all examples of dynamical systems with continuous-time evolution. The two particular examples on spring-mass system and pendulum one are typically simplified (e.g., in textbook approach) so as to include only the first-order interaction. Their corresponding evolution equation is of the form $\frac{d\xbf}{dt} = F(\xbf)$ and $F(\xbf)$ is linear in $\xbf$, while is also time-independent. In this case, such an equation is called a linear dynamical system, and $\xbf(t)$ generally admits a closed-form solution. However, when the condition is more complicated, for example, $F(\xbf)$ involve nonlinear terms in $\xbf$, and may depend on time $t$, then the close-form solution might not exist and understanding the behavior of $\xbf$ becomes challenging. A popular approach to analyze the behavior of state $\xbf$ is employing numerical tools. In fact, tremendous efforts have been made to develop advanced algorithms for solving ODE/PDE, which in turn provide a powerful tool to dissect the properties of dynamical systems.

\subsection{Dynamical system via the scope of TDA }
\label{sec: dynamicalviaTDA}
At the heart of dynamical systems is the evolution equation $\frac{d\xbf}{dt} = F(\xbf,t)$. For a given interval $(0, \tau)$ of interest, as mentioned above, the equation might not be solvable. Thus, knowing $\xbf(t)$ (for $0 \leq t \leq \tau$) is most likely impossible. However, we might be particularly interested in the emerging properties of $\xbf(t)$ within the so-called phase space. More specifically, in this work, we assume that we are working with the Euclidean space $\Rbb^{N}$, and define $\xbf \equiv (x_1,x_2,...,x_N)^T$. The phase space, denoted as $\mathcal{P}$, is then the subspace of $\Rbb^n$ that contains $\xbf(t)$ at different times $t$. By examining the behavior of the trajectory of $\xbf(t)$ in this phase space, we can infer certain physical properties of the original dynamical system. 

To see how TDA can be relevant to the analysis of a dynamical system, we note that within $\mathcal{P}$, $\xbf(t)$ is a data point. Within the interval $(0,\tau)$, we pick $M$ different times $t_1,t_2,..., t_M$, then $\xbf(t_1),\xbf(t_2), ..., \xbf(t_M)$ forms a point cloud. From these points, an edge needs to be assigned between two points. To do this, we need to have a form of distance properly defined between a pair of data points. Although there may be more than one, we note that as these points belong to the Euclidean space $\Rbb^N$, the simplest choice would be the inner product. From the pairwise distances between all pairs of data points, a threshold $\epsilon$ is chosen. Two data points are then connected if their pairwise distance is less than or equal to $\epsilon$. The result of this process is a graph $G=(V,E)$ with $|V| = M$ vertices and corresponding edges that connect the vertices. The simplicial complex is then built from $G$ by identifying a vertex/point as a 0-simplex, an edge as a 1-simplex, a triangle as 2-simplex, and so on.

\begin{figure*}
    \centering
    \begin{tikzpicture}[scale=1.5]
\node[whitev] (T) at (-2.0,2.2) {};
\node (X) at (-2.0, -0.9) {};
\node at (-2.0, -1.0) {$x_1$}; 
\node (Y) at (0.0, 2.2) {};
\node at (0.0, 2.3) {$x_2$};
\node at ( 0.0, 1.5) {A pendulum};
\node at (2.5, 1.5) {2-d topological};
\node at (2.5, 1.3) {sphere};
\draw[<->] (0.2, 1.7) to[out = 90, in = 90] (2.3, 1.7) ;

\node at (-3.6,  0.0) {$t_1$};
\node[yellowv] (A) at (-3.5,0.20) {};
\node at (-2.4, -0.4) {$t_2$};
\node[yellowv] (B) at (-2.4,-0.22) {};
\node at (-1.2,-0.4) {$t_3$};
\node[yellowv] (C) at (-1.2,-0.22) {};
\node at (0.0, 0.2) {$t_4$};
\node[yellowv] (D) at (-0.3,0.20) {};

\draw[->, thick] (T) -- (X);
\draw[->, thick] (T) -- (Y);
\draw[e] (T)--(A);
\draw[e] (T)--(B);
\draw[e] (T)--(C);
\draw[e] (T)--(D);

\draw[->, dashed] (A) -- (B);
\draw[->, dashed] (B) -- (C);
\draw[->, dashed] (C)-- (D);

\draw[->, dashed] (D) to[out = 250, in = -70] (A) ;



\node[yellowv] (P) at (3.7,2.0) {};
\node at (3.7,  2.2) {$\xbf (t_1)$};
\node[yellowv] (Q) at (2.9,0.8) {};
\node at (2.8, 1.0) {$\xbf(t_2)$};
\node[yellowv] (R) at (5.1,1.0) {};
\node at (5.3, 1.2) {$\xbf(t_4)$};
\node[yellowv] (S) at (4.3,-0.2) {};
\node at (4.3, -0.4) {$\xbf(t_3)$};

\draw[e] (P)--(Q);
\draw[e] (P)--(R);
\draw[e] (R)--(S);
\draw[e] (Q) -- (S);
\end{tikzpicture}
    \caption{A simple illustration of how TDA can ``view'' the dynamical system. (\textbf{Left figure}) A simple pendulum is composed of a mass $M$ connected to some fixed point by a string (assuming of negligible mass). Provided an initialization, for example, shifting the mass so as to create an angle $\theta_0$. Under the influence of gravity, the pendulum will be swinging back and forth, creating a periodic motion, e.g., the trajectory is indicated by the dashed line, at different times $t_1,t_2,t_3,t_4$. In fact, in this case, the coordinate $(x_1,x_2)$ of the mass can be written down explicitly by solving the Newton's equation of motion. (\textbf{Right figure}) Within the TDA framework, by defining $\xbf \equiv (x_1,x_2)$ and treat $\xbf(t_1), \xbf(t_2), \xbf(t_3), \xbf(t_4)$ as data points/vectors in $\Rbb^2$. The distance between two data points is given by the inner product between them. Two points are connected if their distance are less than a chosen threshold $\epsilon$.   }
    \label{fig: illustration}
\end{figure*}
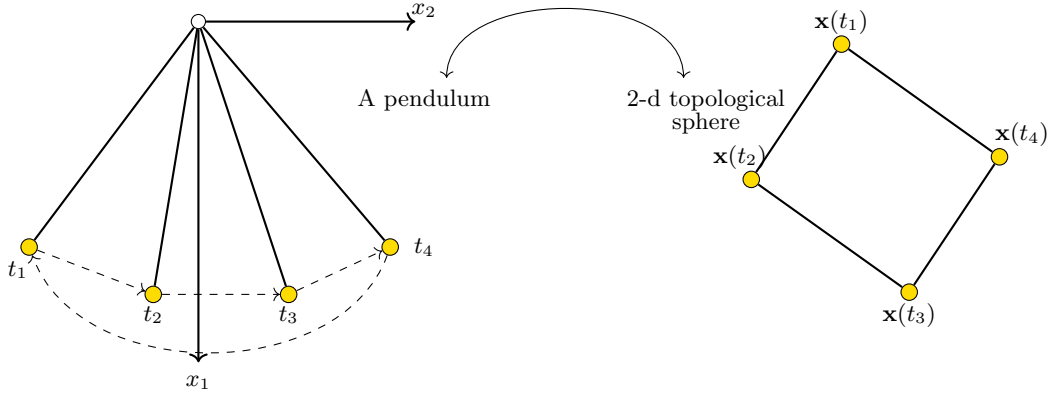

The above procedure shows how a dynamical system can fit into the TDA framework. The next step is to elaborate on how the outcome of such a TDA framework can provide insight toward the original dynamical system. We recall earlier that for a given complex, the Betti numbers capture the number of ``holes''. The first Betti number $\beta_1$ indicates the number of loops. If the value of $\beta_1$ is non-zero, it implies that the trajectory of $\xbf$ in the phase space is tentatively periodic. Furthermore, in the previous paragraph, we mentioned the threshold $\epsilon$ which is chosen to assign an edge to the complex. In effect, the value of $\epsilon$ controls the density of the edge in the graph. The higher the value of $\epsilon$, the more edges would enter the graph, and thus the complex would contain more simplex. If the values of Betti numbers over a range of threshold vary significantly, it can be a signal of chaotic. As an example, we provide the Fig.~\ref{fig: illustration} to show how a simple dynamical system, e.g., pendulum, is viewed under the perspective of TDA. In this particular case, the calculation of Betti numbers results in the first Betti number, $\beta_1 = 1$, and the remaining are zero, i.e., a 1-d topological sphere. This reflects the periodicity of the original pendulum system, which is known to have such a property.

We note that topological and geometrical insight has been applied to understanding dynamical systems in a variety of works \cite{packard1980geometry, noakes1991takens, perea2015sliding, de2009persistent, kaczynski2006computational,batko2020conley,khasawneh2016chatter, myers2019persistent}. Aside from the example provided in Fig.~\ref{fig: illustration}, more intrinsic properties of dynamical systems can be inferred from the topological information. The work in \cite{perea2015sliding}, for example, has revealed that the quasi-periodic behavior of the trajectory (of some dynamical system) can be reflected via a torus-like structure of the corresponding simplcial complex. 

As the next step, we shall proceed to describe our main proposal, which is a quantum TDA framework for probing dynamical systems. 

\section{Quantum topological data analysis algorithm for dynamical system}
\label{sec: qTDAfordynamicalsystem}
Before describing our main proposal, we provide an overview of existing quantum algorithms for ODE and TDA. 

\subsection{An overview of existing quantum algorithms for ODE}
\label{sec: overviewODE}
As mentioned above, dynamical systems, especially nonlinear systems, exhibit nontrivial challenge for theoretical analysis. Although numerical approaches have been proposed to deal with this case, another type of challenge emerges with respect to computational resources. Typically, a dynamical system is discretized, for example, time interval $(0,\tau)$ (for ODE) or domain interval (for PDE), into smaller components. The number of components can be very substantial, even exponentially large with respect to the size of input system. Thus, it is very demanding to store and process these amounts of components. 

Quantum computing solutions to the above issues have been proposed in a variety of works \cite{leyton2008quantum, liu2021efficient, zanger2021quantum, berry2014high, bagherimehrab2023fast, krovi2023improved, liu2023efficient, childs2020quantum, jin2024quantum, childs2021high, linden2022quantum, tanaka2023polynomial, shang2024design, tennie2025quantum, jin2025schrodingerization}, tailored to handle many different kinds of ODE. For example, the work \cite{leyton2008quantum} considers a system of ODE of the form:
\begin{align}
    \begin{cases}
        \frac{d x_1(t)}{dt} = f_1(x_1,x_2,..., x_n,t) \\
        \vdots \\
        \frac{d x_2(t)}{dt} = f_n(x_1,x_2,..., x_n,t) 
    \end{cases}
\end{align}
where each $f_i(x_1,x_2,...,x_n,t)$ is a polynomial in the variables $x_1,x_2,...,x_n,t$. For convenience and subsequent use, we define $\xbf \equiv  (x_1,x_2,...,x_n)^T$ and $F \equiv (f_1,f_2,...,f_n)^T $. Their method relies on two parts. First, encode the variables $\{x_i\}_{i=1}^n$ into the amplitudes of a quantum state $\sum_{i=1}^n x_i \ket{i-1}$. Second, break the time interval $[0,\tau]$ into equal subintervals $[0,t_1,t_2,..., t_N \equiv \tau]$ and define $\Delta \equiv t_{i+1}- t_i$, then using forward Euler method to integrate the system as:
\begin{align}
    x_i( t_{j+1}) = x_i(t_j) + \frac{d x_i(t)}{dt} \Delta = x_i(t_j) + f_i(t_j) \Delta 
\end{align}
which results in integration of the entire system $\xbf(t_{j+1}) = \xbf(t_j) + F \Delta $. By repeating the above procedure $N$ times, the outcome is a quantum state $\varpropto \sum_{i=1}^n x_i(\tau) \ket{i-1}$ corresponding to the solution at time $\tau$ of interest. 

The work in \cite{berry2017quantum} considers a system of $n$-variables ODE of the form:
\begin{align}
    \frac{d \xbf}{dt} = A \xbf +  \textbf{b}
\end{align}
where $A \in \Rbb^{n \times n}, \textbf{b} \in \Rbb^n$ are time-independent. As indicated in \cite{berry2017quantum}, the system above has an exact solution:
\begin{align}
    \xbf(t) = \exp(A t) \xbf(0) + (\exp(At) - \Ibb) A^{-1} \textbf{b}
\end{align}
Another properties they use are the approximation to the exponential function:
\begin{align}
    \exp(z) \approx \sum_{j=0}^k \frac{z^j}{j!} \\
    (\exp(z) - 1) z^{-1} \approx \sum_{j=1}^k \frac{z^{j-1}}{j!}
\end{align}
At time $t_1$, the solution is:
\begin{align}
    \xbf(t_1) \approx \exp( A t_1) \xbf(0) + ( \exp( A t_1) - \Ibb) h \textbf{b}
\end{align}
Similarly, at $t_2$, we treat $\xbf(t_1)$ as an input: 
\begin{align}
    \xbf(t_2) = \exp(A t_2) \xbf(t_1) + (\exp(A t_2) - \Ibb) h \textbf{b}
\end{align}
Repeating the same procedure for different times $t_3,...,t_N \equiv \tau$, a linear system can be obtained. The solution to such linear system is $\varpropto \sum_{i=1}^N \xbf(t_i) \ket{i-1}$, and can be obtained via quantum linear solving algorithms \cite{harrow2009quantum, childs2017quantum}.

The work in \cite{berry2014high} focuses on the following type of ODE:
\begin{align}
    \frac{d \xbf(t)}{dt} = A(t) \xbf(t) + \textbf{b}(t)
\end{align}
where $\xbf(t) \in \Rbb^n$, $A(t) \in \Rbb^{n \times n}$ and $\textbf{b}(t) \in \Rbb^n$. The strategy proposed in \cite{berry2014high} is to discretize the above system, in order to apply the quantum linear solving algorithm \cite{harrow2009quantum, childs2017quantum}. More concretely, break the time interval $[0,\tau]$ into $[0,t_1,t_2,..., t_N \equiv \tau]$ as above, then the derivative is approximated as follows:
\begin{align}
   \frac{d \xbf(t_i)}{dt} = \frac{\xbf(t_{i+1}) - \xbf(t_i)  )}{ \Delta}
\end{align}
which can be plugged into the ODE to obtain:
\begin{align}
    \frac{\xbf(t_{i+1}) - \xbf(t_i)  )}{ \Delta} = A(t_i) \xbf(t_i) + \textbf{b}(t_i)
\end{align}
Repetition of the same algebraic procedure yields a linear system where the solution of such a linear equation contains the values of $\xbf(t)$ over times in the interval $[0,\tau]$. From here, one can leverage the quantum algorithm for solving linear equations \cite{harrow2009quantum, childs2017quantum} to obtain a quantum state that is $\propto \sum_{i=1}^N \ket{i} \xbf(t_i)$. 

The work in \cite{childs2020quantum} also considers the same type of ODE as above; however, they use different method to solve. Instead of discretizing the system to directly produce the linear system that contains the solution to the given ODE, Ref.~\cite{childs2020quantum} builds on the spectral method. This method approximates the solution at a given time $t$ as:
\begin{align}
    \xbf(t) = \sum_{k=1}^N \textbf{c}_k T_k(t)
\end{align}
where $\textbf{c}_k \in \Rbb^n$ is $n$-dimensional vector and $T_k(t)$ is the Chebyshev polynomial. As mentioned in \cite{childs2020quantum}, the coefficients of $\{ \textbf{c}_k \}$ can be found by requiring $\xbf(t)$ to satisfy the ODE and the initial conditions at a set of the namely interpolation nodes. From this, one can construct a linear equation with the entries of $\{ \textbf{c}_k\}_{k=1}^N$ as main variables. Solving such a linear equation produces $\{\textbf{c}_k\}$ accordingly, and the quantum state $\ket{\xbf(t)} \varpropto \xbf(t)$ can be obtained. 

We remark that there is a rich body of literature in this direction. The examples above cover only a few of them. As analyzed in these aforementioned works, the promise of quantum algorithms rely on achieving polylogarithmic scaling in the size of inputs. However, the output of these quantum ODE solvers is typically a quantum state $\ket{\xbf(\tau)}$ corresponding to $\xbf$ at the desired time $\tau$. If we want to know such a solution explicitly, then we need to perform tomography, which can be very costly and void all the quantum speedup. An open aspect in the area of quantum algorithms has been:
\noindent
\begin{center}
    \textbf{Open aspect:} Can the quantum solution state $\ket{\xbf(\tau)}$ be leveraged for useful purposes ?
\end{center}
This aspect has also been mentioned in \cite{harrow2009quantum} and in the same work, the authors suggest that instead of finding explicit solution to the linear system $A\xbf = b$, one is interested in some expectation value $\bra{\xbf} M \ket{\xbf}$. In a similar manner, the work \cite{clader2013preconditioned} also suggests that from the quantum state representation, one can use it to compute electromagnetic cross-area. A more thorough and critical discussion regarding this quantum output issue can be found in \cite{aaronson2015read}. 

Below, we proceed to provide an overview of quantum TDA, before we describe our main proposal for combining quantum TDA algorithms and quantum ODE solvers. It will be shown that the quantum output of the quantum ODE solvers can be integrated with the quantum TDA algorithm, thus forming a framework for probing the dynamical system. In turn, it provides an affirmative answer, adding further convincing example for the open aspect above.

\subsection{An overview of existing quantum algorithms for TDA}
\label{sec: overviewquantumTDA}
The first quantum algorithm for estimating Betti numbers -- a central problem within TDA -- was introduced in \cite{lloyd2016quantum}. Their algorithm is built on a few well-known quantum primitives, including the quantum search algorithm \cite{grover1996fast}, the Hamiltonian simulation \cite{berry2007efficient,berry2012black,berry2015hamiltonian} and the quantum phase estimation algorithm \cite{kitaev1995quantum}. Subsequently, the works in \cite{ubaru2021quantum} and \cite{berry2024analyzing} provide a certain improvement over \cite{lloyd2016quantum} in many aspects. For example, by exploiting the connection between the boundary operator and the fermionic operator, the authors in \cite{ubaru2021quantum} are able to replace the Hamiltonian simulation subroutine in \cite{lloyd2016quantum} with a faster procedure, resulting in a more efficient algorithm. Another important part of \cite{ubaru2021quantum} is the application of stochastic rank estimation method \cite{ubaru2016fast,ubaru2017fast} to replace the quantum phase estimation subroutine in \cite{lloyd2016quantum}, which significantly reduce the quantum circuit depth. The work in \cite{berry2024analyzing}, at the same time, introduces new method for state preparation, e.g., Dicke state preparation via inequality testing, and a more efficient method for amplitude estimation via the namely Kaiser window.

In a few parallel attempts, it has been shown that estimating Betti numbers is not trivial. Specifically, the work \cite{schmidhuber2022complexity} shows that such estimation is $\rm NP$-hard, and the work \cite{crichigno2024clique} shows that it is $\rm QMA$-hard. These complexity-theoretic results have implied that in most cases, quantum algorithms only yield at most polynomial speedup. Recently, the works \cite{lee2025new, nghiem2025hybrid, nghiem2023quantum} explore a few alternatives and show that exponential speedup in estimating Betti numbers is still possible in structured settings. The works \cite{hayakawa2022quantum, gyurik2024quantum}, on the other hand, consider the problem of estimating persistent Betti numbers, which is a generalization of Betti numbers. In particular, the work \cite{gyurik2020towards} have shown that estimating the namely normalized Betti numbers can be difficult for classical computers, while being easy for quantum one. 

The above paragraphs showcase a brief survey of existing literature in quantum TDA. We keep things neat for simplicity, and refer the curious readers to these cited works for more details. A more technical overview of quantum TDA can be found in Section II.A of \cite{berry2024analyzing}.

\subsection{Our proposal}
\label{sec: mainproposal}
As can be seen from Section \ref{sec: overviewODE}, the output of those quantum ODE solvers is typically a quantum state $\ket{\xbf}$ that is proportional to $\xbf$ at time $\tau$. Some algorithms, for example, \cite{childs2020quantum}, output a quantum state of the form $\sum_{i=1}^N \ket{i} \xbf( t_i)$, with $t_i$ being the time within $ (0, \tau)$, e.g., dividing $(0,\tau) $ into $ (0,t_1) \cup (t_1,t_2)  \cup ... \cup (t_i,t_{i+1}), \cup ..., \cup  (t_{N-1}, t_N \equiv \tau) $. To obtain the state $\ket{\xbf(t_i)}$, we can measure $\sum_{i=1}^N \ket{i} \xbf( t_i)$ and post-select the measurement outcome $\ket{i}$. 

In practice, depending on the dynamical system of interest, we would need to select the appropriate quantum algorithm among the aforementioned works, for the corresponding ODE. In the following, we proceed with the premise that, as an output of the algorithms above, we have the quantum states at different times $\ket{\xbf(t_1)}, \ket{\xbf (t_2)}, ..., \ket{\xbf (t_M)}$. We then build a graph $G$ as follows.

\noindent
\textbf{Building graph $G$ associated with the given dynamical system: }
\begin{itemize}
    \item Associate $M$ vertices $v_0,v_1,...,v_{M-1}$.
    \item We then use either the SWAP/Hadamard test to estimate the inner product between a pair of quantum states $d_{ij} \equiv |\braket{\xbf_i,\xbf_j}|$. 
    \item Choose a threshold $\epsilon$. 
    \item Connect two vertices $v_i,v_j$ if $d_{ij} \leq \epsilon$.
\end{itemize}
From the classical knowledge of $G$, we can apply existing quantum TDA algorithms to analyze its topological structure, particularly \cite{nghiem2025hybrid} and \cite{berry2024analyzing}. The reason for these choices is that, the hybrid framework in \cite{nghiem2025hybrid} admits the classical knowledge of graph $G$ as input. The framework in \cite{berry2024analyzing} is fully quantum, however, a key component in this algorithm, which is the oracle that verifies the existence of simplexes, can be explicitly constructed from the graph knowledge, i.e., the list of vertices and edges. In addition, this framework is also the best up-to-date quantum algorithm for estimating Betti numbers, building upon the initial LGZ algorithm \cite{lloyd2016quantum}.

To elaborate on these results, we mention that the method in \cite{nghiem2025hybrid} is a hybrid classical-quantum algorithm. From the graph's knowledge (the list of vertices and edges), one can leverage the classical algorithm in \cite{chiba1985arboricity, eppstein2010listing} to enumerate the simplexes. Such knowledge of simplexes is then used as input to the quantum algorithm developed in \cite{lee2025new}, so as to estimate Betti numbers. For our purpose, we recapitulate the result of \cite{nghiem2025hybrid} in the following lemma, and refer to \cite{nghiem2025hybrid} for more details. 
\begin{lemma}[\cite{nghiem2025hybrid}]
\label{lemma: hybrid}
    Let the classical knowledge of the graph $G = (V,E)$ corresponding to the simplex $K$ of interest be given. Let $|V| = n$ be the total number of data points and $|E| $ is the total number of edges in $G$ (also $K$). Then with success probability $\geq 1-\eta$, the $r$-th normalized Betti number $\frac{\beta_r}{|S_r|}$ can be estimated to an additive accuracy $\epsilon$ using:
    \begin{itemize}
       \item  A classical algorithm with complexity
    \begin{equation}
        \mathcal{O}\left( |E| \alpha(\mathcal{G})^{r-1} \right) \quad \text{or} \quad \mathcal{O}\left( dn \cdot 3^{d/3} \right),
    \end{equation}
    where $\alpha(\mathcal{G})$ and $d$ denote the arboricity and degeneracy of graph $\mathcal{G}$, respectively. These two numbers are upper bounded by $\max_{v \in V} \deg(v)$.
    \item A quantum circuit of depth
    \begin{equation}
        \mathcal{O}\left( n + \log |S_r|  \right)
    \end{equation}
    executed $\mathcal{O}\left(\log\left( r|S_r| |S_{r+1}|\right) \log\left(\frac{1}{\eta}\right)\frac{1}{\epsilon^2}\right)$ times, using a total of $\mathcal{O}\left( n + \log |S_r| \right)$ qubits.
    \end{itemize}
\end{lemma}

The method in \cite{berry2024analyzing} is based on the LGZ algorithm, with some improvements. First, they encode the simplex into the computational basis in a similar manner to \cite{lloyd2016quantum}. Then they aim to block-encode the namely Dirac operator $B_G$, which is defined as:
\begin{align}
    B_G = \begin{pmatrix}
        0 &  \partial_{r-1} & 0 \\
        \partial_{r-1}^\dagger & 0 & \partial_r \\
        0 & \partial_r^\dagger & 0 
    \end{pmatrix}
\end{align}
Then the (normalized) Betti numbers can be found by projecting the above (block-encoded) operator onto the zero eigenspace, followed by an amplitude estimation step. For our purpose, we summarize the main result of \cite{berry2024analyzing} as follows. 
\begin{lemma}[\cite{berry2024analyzing}]
\label{lemma: quantum}
Let the classical knowledge of the graph $G = (V,E)$ corresponding to the simplex $K$ of interest be given. Then with a success probability $\geq 1-\eta$, the $r$-th normalized Betti number $\frac{\beta_r}{|S_r|}$ can be estimated to an additive error $\epsilon$ using a quantum circuit of depth 
\begin{align}
    6 |E| \ln \left(  \frac{1}{\eta} \right) \frac{1}{\sqrt{\epsilon}} \left( \frac{\pi}{2}  \sqrt{\frac{\binom{n}{r}}{ |S_r|}} +  \frac{n}{\lambda_{\min} } \log \frac{1}{\epsilon}\right)
\end{align}
\end{lemma}

Earlier we have discussed how dynamical systems are viewed under the scope of TDA, where the appearance of ``holes'' can be an indication of periodicity, quasi-periodicity, etc. Essentially, Betti numbers can capture the number of ``holes''. The outputs of the above two lemmas are actually normalized Betti numbers. As also mentioned in previous works, e.g., \cite{schmidhuber2022complexity}, an $\epsilon$ additive error estimation of $\beta_r/|S_r|$ can be translated to an $\delta$ multiplicative precision of $\beta_r$ by choosing $\epsilon = \delta \frac{\beta_r}{|S_r|}$. As a result, the complexity will be proportional to $\frac{|S_r|^2}{\beta_r^2}$ (in Lemma \ref{lemma: hybrid}) or $ \sqrt{\frac{|S_r|}{\beta_r}}$ (in Lemma \ref{lemma: quantum}). This is most efficient when the given complex has large Betti numbers, i.e., $\beta_r \approx |S_r|$. A family of graphs/complexes that have this property can be found in \cite{berry2024analyzing}.

\section{Conclusion}
\label{sec: conclusion} 
In this work, we have proposed a quantum framework to analyze dynamical systems. We particularly combine the ideas of  quantum algorithm for ordinary differential equations and topological data analysis. We show that the output of quantum ODE solvers, which are typically quantum states, can be leveraged to obtain the classical knowledge of a graph. The knowledge of such a graph is then used as input to two quantum TDA algorithms, specifically \cite{nghiem2025hybrid} and \cite{berry2024analyzing}. These two algorithms then return the estimation of (normalized) Betti numbers, and we have elaborated on how these numbers -- which quantify the number of ``holes'' in the corresponding simplicial complex, can be used to deduce certain dynamical properties. We believe that our work can serve as a first step toward the application of quantum ODE solvers and of quantum TDA algorithms as well.

\section*{Acknowledgements}
We thank Jordan Cotler for interesting discussion that resulted in the project. Part of this work is done when the author is an intern at QuEra Computing Inc. We particularly acknowledge the hospitality of Harvard University where the author has an academic visit during the completion of this project.

\bibliography{ref.bib}
\bibliographystyle{unsrt}

\newpage
\appendix
\onecolumngrid

\section{Block-encoding and quantum singular value transformation}
\label{sec: summaryofnecessarytechniques}
We briefly summarize the essential quantum tools used in our algorithm. For conciseness, we highlight only the main results and omit technical details, which are thoroughly covered in~\cite{gilyen2019quantum}. An identical summary is also presented in~\cite{lee2025new}.

\begin{definition}[Block-encoding unitary, see e.g.~\cite{low2017optimal, low2019hamiltonian, gilyen2019quantum}]
\label{def: blockencode} 
Let $A$ be a Hermitian matrix of size $N \times N$ with operator norm $\norm{A} < 1$. A unitary matrix $U$ is said to be an \emph{exact block encoding} of $A$ if
\begin{align}
    U = \begin{pmatrix}
       A & * \\
       * & * \\
    \end{pmatrix},
\end{align}
where the top-left block of $U$ corresponds to $A$. Equivalently, one can write
\begin{equation}
    U = \ket{\mathbf{0}}\bra{\mathbf{0}} \otimes A + (\cdots),    
\end{equation}
where $\ket{\mathbf{0}}$ denotes an ancillary state used for block encoding, and $(\cdots)$ represents the remaining components orthogonal to $\ket{\mathbf{0}}\bra{\mathbf{0}} \otimes A$. If instead $U$ satisfies
\begin{equation}
    U = \ket{\mathbf{0}}\bra{\mathbf{0}} \otimes \tilde{A} + (\cdots),
\end{equation}
for some $\tilde{A}$ such that $\|\tilde{A} - A\| \leq \epsilon$, then $U$ is called an {$\epsilon$-approximate block encoding} of $A$. Furthermore, the action of $U$ on a state $\ket{\mathbf{0}}\ket{\phi}$ is given by
\begin{align}
    \label{eqn: action}
    U \ket{\mathbf{0}}\ket{\phi} = \ket{\mathbf{0}} A\ket{\phi} + \ket{\mathrm{garbage}},
\end{align}
where $\ket{\mathrm{garbage}}$ is a state orthogonal to $\ket{\mathbf{0}}A\ket{\phi}$. The circuit complexity (e.g., depth) of $U$ is referred to as the {complexity of block encoding $A$}.
\end{definition}

Based on~\cref{def: blockencode}, several properties, though immediate, are of particular importance and are listed below.
\begin{remark}[Properties of block-encoding unitary]
The block-encoding framework has the following immediate consequences:
\begin{enumerate}[label=(\roman*)]
    \item Any unitary $U$ is trivially an exact block encoding of itself.
    \item If $U$ is a block encoding of $A$, then so is $\Ibb_m \otimes U$ for any $m \geq 1$.
    \item The identity matrix $\Ibb_m$ can be trivially block encoded, for example, by $\sigma_z \otimes \Ibb_m$.
\end{enumerate}
\end{remark}

Given a set of block-encoded operators, various arithmetic operations can be done with them. Here, we simply introduce some key operations that are especially relevant to our algorithm, focusing on how they are implemented and their time complexity, without going into proofs. For more detailed explanations, see~\cite{gilyen2019quantum, camps2020approximate}.

\begin{lemma}[Informal, product of block-encoded operators, see e.g.~\cite{gilyen2019quantum}]
\label{lemma: product}
    Given unitary block encodings of two matrices $A_1$ and $A_2$, with respective implementation complexities $T_1$ and $T_2$, there exists an efficient procedure for constructing a unitary block encoding of the product $A_1 A_2$ with complexity $T_1 + T_2$.
\end{lemma}

\begin{lemma}[Informal, tensor product of block-encoded operators, see e.g.~{\cite[Theorem 1]{camps2020approximate}}]\label{lemma: tensorproduct}
    Given unitary block-encodings $\{U_i\}_{i=1}^m$ of multiple operators $\{M_i\}_{i=1}^m$ (assumed to be exact), there exists a procedure that constructs a unitary block-encoding of $\bigotimes_{i=1}^m M_i$ using a single application of each $U_i$ and $\mathcal{O}(1)$ SWAP gates.
\end{lemma}

\begin{lemma}[Informal, linear combination of block-encoded operators, see e.g.~{\cite[Theorem 52]{gilyen2019quantum}}]
    Given the unitary block encoding of multiple operators $\{A_i\}_{i=1}^m$. Then, there is a procedure that produces a unitary block encoding operator of $\sum_{i=1}^m \pm (A_i/m) $ in time complexity $\mathcal{O}(m)$, e.g., using the block encoding of each operator $A_i$ a single time. 
    \label{lemma: sumencoding}
\end{lemma}

\begin{lemma}[Informal, Scaling multiplication of block-encoded operators] 
\label{lemma: scale}
    Given a block encoding of some matrix $A$, as in~\cref{def: blockencode}, the block encoding of $A/p$ where $p > 1$ can be prepared with an extra $\mathcal{O}(1)$ cost.
\end{lemma}



\begin{lemma}[Matrix inversion, see e.g.~\cite{gilyen2019quantum, childs2017quantum}]\label{lemma: matrixinversion}
Given a block encoding of some matrix $A$  with operator norm $||A|| \leq 1$ and block-encoding complexity $T_A$, then there is a quantum circuit producing an $\epsilon$-approximated block encoding of ${A^{-1}}/{\kappa}$ where $\kappa$ is the conditional number of $A$. The complexity of this quantum circuit is $\mathcal{O}\left( \kappa T_A \log \left({1}/{\epsilon}\right)\right)$. 
\end{lemma}

\begin{lemma}\label{lemma: amp_amp}[\cite{gilyen2019quantum} Theorem 30]
\label{lemma: amplification}
Let $U$, $\Pi$, $\widetilde{\Pi} \in {\rm End}(\mathcal{H}_U)$ be linear operators on $\mathcal{H}_U$ such that $U$ is a unitary, and $\Pi$, $\widetilde{\Pi}$ are orthogonal projectors. 
Let $\gamma>1$ and $\delta,\epsilon \in (0,\frac{1}{2})$. 
Suppose that $\widetilde{\Pi}U\Pi=W \Sigma V^\dagger=\sum_{i}\varsigma_i\ket{w_i}\bra{v_i}$ is a singular value decomposition. 
Then there is an $m= \mathcal{O} \Big(\frac{\gamma}{\delta}
\log \left(\frac{\gamma}{\epsilon} \right)\Big)$ and an efficiently computable $\Phi\in\mathbb{R}^m$ such that
\begin{equation}
\left(\bra{+}\otimes\widetilde{\Pi}_{\leq\frac{1-\delta}{\gamma}}\right)U_\Phi \left(\ket{+}\otimes\Pi_{\leq\frac{1-\delta}{\gamma}}\right)=\sum_{i\colon\varsigma_i\leq \frac{1-\delta}{\gamma} }\tilde{\varsigma}_i\ket{w_i}\bra{v_i} , \text{ where } \Big|\!\Big|\frac{\tilde{\varsigma}_i}{\gamma\varsigma_i}-1 \Big|\!\Big|\leq \epsilon.
\end{equation}
Moreover, $U_\Phi$ can be implemented using a single ancilla qubit with $m$ uses of $U$ and $U^\dagger$, $m$ uses of C$_\Pi$NOT and $m$ uses of C$_{\widetilde{\Pi}}$NOT gates and $m$ single qubit gates.
Here,
\begin{itemize}
\item C$_\Pi$NOT$:=X \otimes \Pi + I \otimes (I - \Pi)$ and a similar definition for C$_{\widetilde{\Pi}}$NOT; see Definition 2 in \cite{gilyen2019quantum},
\item $U_\Phi$: alternating phase modulation sequence; see Definition 15 in \cite{gilyen2019quantum},
\item $\Pi_{\leq \delta}$, $\widetilde{\Pi}_{\leq \delta}$: singular value threshold projectors; see Definition 24 in \cite{gilyen2019quantum}.
\end{itemize}
\end{lemma}

The above lemmas are the basic recipes involving block-encoding operators, which were originally derived in \cite{gilyen2019quantum}. The following ones are slightly more advanced, appearing in the recent works \cite{nghiem2023quantum, lee2025new}.
\begin{lemma}[\cite{nghiem2025refined}]
\label{lemma: entrycomputablematrix}
    Let $A$ be an $M \times N$ matrix with classically known entries and Frobenius norm $\norm{A}_F$. Then there exists a quantum circuit of depth $\mathcal{O}\left(\log(s N)\right)$ that implements a block-encoding of $A^\dagger A / \norm{A}_F^2$.
    \end{lemma}
For the readers who are interested in the proof, see Appendix S2 of \cite{nghiem2025refined}. We remark an important point that the procedure underlying the above lemma actually employs the state preparation technique in \cite{zhang2022quantum}. This preparation step requires a classical pre-processing of complexity $\mathcal{O}(\log N)$. However, such complexity can be reduced to $\mathcal{O}(1)$, as in this case, the boundary operator $\partial_r$ has entries to be either $1$ or $-1$, and thus the classical pre-processing cost is $\mathcal{O}(1)$. 

\begin{lemma}[\cite{nghiem2023quantum}]
\label{lemma: removingfactor}
    Let $U_A$ be an exact block-encoding of an operator $A/\alpha$, with operator norm of $A/\alpha$, $||A||/\alpha \leq 1$ and $\alpha \geq 1$. Then there is a quantum circuit of depth $\mathcal{O}\left(  \kappa^2 T_A \log^4 \frac{\alpha}{\epsilon} \right) $ that implements the $\epsilon$-approximated block-encoding of $A/\kappa$, where $\kappa$ is the condition number of $A$ and $T_A$ is the circuit complexity of $U_A$. 
\end{lemma}
We refer the interested readers to the Appendix F of \cite{nghiem2023quantum} for the proof. Here, we point out and discuss the importance of the above lemma. It can be seen that if $\kappa \leq \alpha$, then to obtain the block-encoding of $A/\kappa$, we can use Lemma \ref{lemma: amp_amp} to multiply the block-encoding of $A/\alpha$ with $\alpha/\kappa$. This way has complexity $\mathcal{O}( \alpha/\kappa)$ which is linear in $\alpha$. At the same time, the above lemma achieves better dependence on $\alpha$, with a trade-off being the dependence on condition number $\kappa$. As we see throughout our work, the factor (similar to $\alpha$) associated with the block-encoding of the boundary operator is usually very high (like of the order of dimension of corresponding matrix), thus the above lemma allows a more efficient way to remove it.

\section{A Review of Algebraic Topology}
\label{sec: reviewofalgebraictopology}

This appendix provides a concise overview of the fundamental concepts in algebraic topology, which is based on \cite{nakahara2018geometry}. An identical summary can be found in \cite{lee2025new, nghiem2025hybrid}.

\begin{definition}[Simplex]
\label{def:simplex}
Let $p_0, p_1, \ldots, p_r \in \mathbb{R}^m$ be geometrically independent points where $m \geq r$. The $r$-simplex $\sigma_r = [p_0, p_1, \ldots, p_r]$ is defined as:
\begin{equation}
\sigma_r = \left\{ x \in \mathbb{R}^m : x = \sum_{i=0}^r c_i p_i, \quad c_i \geq 0, \quad \sum_{i=0}^r c_i = 1 \right\},
\end{equation}
where the coefficients $(c_0, c_1, \ldots, c_r)$ are called the barycentric coordinates of $x$.
\end{definition}

Geometrically, a 0-simplex $[p_0]$ represents a point, a 1-simplex $[p_0, p_1]$ represents a line segment, a 2-simplex $[p_0, p_1, p_2]$ represents a triangle, and higher-dimensional simplices generalize this pattern to higher dimensions.

An $r$-simplex can be assigned an orientation. For instance, the 1-simplex $[p_0, p_1]$ has orientation $p_0 \to p_1$, which differs from $[p_1, p_0]$. Throughout this work, we adopt the convention that for an $r$-simplex $[p_0, p_1, \ldots, p_r]$, the indices are ordered from low to high, indicating the canonical orientation.

\begin{definition}[Face and simplicial complex]
For an $r$-simplex $[p_0, p_1, \ldots, p_r]$, any $(s+1)$-subset of its vertices defines an $s$-face $\sigma_s$ where $s \leq r$. A simplicial complex $K$ is a finite collection of simplices satisfying:
\begin{enumerate}
\item Every face of a simplex in $K$ is also in $K$
\item The intersection of any two simplices in $K$ is either empty or a common face of both simplices
\end{enumerate}
The dimension of $K$ is $\dim(K) = \max\{r : \sigma_r \in K\}$.
\end{definition}

\begin{definition}[Chain group]
\label{def: chaingroup}
Let $K$ be an $n$-dimensional simplicial complex. The $r$-th chain group $C_r^K$ is the free abelian group generated by the oriented $r$-simplices of $K$. For $r > \dim(K)$, we define $C_r^K = 0$. Formally, let $S_r^K = \{\sigma_{r,1}, \sigma_{r,2}, \ldots, \sigma_{r,|S_r^K|}\}$ denote the set of $r$-simplices in $K$. An $r$-chain is an element of the form:
\begin{equation}
c_r = \sum_{i=1}^{|S_r^K|} c_i \sigma_{r,i}
\end{equation}
where $c_i \in \mathbb{R}$ are real coefficients. The group operation is defined by:
\begin{equation}
c_r^{(1)} + c_r^{(2)} = \sum_{i=1}^{|S_r^K|} \left( c_i^{(1)} + c_i^{(2)} \right) \sigma_{r,i},
\end{equation}
making $C_r^K$ a free abelian group of rank $|S_r^K|$.
\end{definition}
In fact, since the coefficients $\{ c_i \}$ belong to a field $\Rbb$, the group $C_r^K$ is also a vector space. In the following, we use the Abelian group/vector space interchangeably. The boundary operator $\partial_r : C_r^K \to C_{r-1}^K$ is fundamental to homological algebra.

\begin{definition}[Boundary operator]
For an $r$-simplex $[p_0, p_1, \ldots, p_r]$, the boundary operator is defined as:
\begin{equation}
\partial_r [p_0, p_1, \ldots, p_r] = \sum_{i=0}^r (-1)^i [p_0, p_1, \ldots, \hat{p_i}, \ldots, p_r],
\end{equation}
where $\hat{p_i}$ indicates that vertex $p_i$ is omitted, yielding an $(r{-}1)$-simplex.

For an $r$-chain $c_r = \sum_{i=1}^{|S_r^K|} c_i \sigma_{r,i}$, we extend linearly:
\begin{equation}
\partial_r c_r = \sum_{i=1}^{|S_r^K|} c_i \partial_r \sigma_{r,i}.
\end{equation}
\end{definition}

The boundary operators form a chain complex:
\begin{equation}
0 \xrightarrow{} C_n^K \xrightarrow{\partial_n} C_{n-1}^K \xrightarrow{\partial_{n-1}} \cdots \xrightarrow{\partial_1} C_0^K \xrightarrow{\partial_0} 0.
\end{equation}

\begin{definition}[Cycles, boundaries, and homology]
An $r$-chain $c_r$ is called an $r$-cycle if $\partial_r c_r = 0$. The collection of all $r$-cycles forms the $r$-cycle group $Z_r^K = \ker(\partial_r)$. Conversely, an $r$-chain $c_r$ is called an $r$-boundary if there exists an $(r{+}1)$-chain $d_{r+1}$ such that $\partial_{r+1} d_{r+1} = c_r$. The set of all $r$-boundaries forms the $r$-boundary group $B_r^K = \textnormal{im}(\partial_{r+1})$.
\end{definition}
A fundamental property of boundary operators is that $\partial_r \circ \partial_{r+1} = 0$, which ensures that every boundary is also a cycle, i.e., $B_r^K \subseteq Z_r^K$. This inclusion allows us to define the $r$-th homology group/space as the quotient group/space $H_r^K = Z_r^K / B_r^K$, which captures the notion of cycles that are not boundaries.

\begin{definition}[Betti numbers]
The $r$-th Betti number is defined as:
\begin{equation}
\beta_r = \textnormal{dim}(H_r^K) = \textnormal{dim}(Z_r^K) - \textnormal{dim}(B_r^K).
\end{equation}
\end{definition}
In the group-theoretic language, the above dimension, e.g., $\textnormal{dim}(H_r^K) $, is replaced by $\rm rank (H_r^K)$. As we mentioned, we use the notion of group/space interchangeably. For computational purposes, we can utilize the combinatorial Laplacian:
\begin{equation}
\Delta_r = \partial_{r+1} \partial_{r+1}^\dagger + \partial_r^\dagger \partial_r,
\end{equation}
where $\partial_r^\dagger$ denotes the adjoint of $\partial_r$. A fundamental result in algebraic topology establishes that:
\begin{equation}
H_r^K \cong \ker(\Delta_r),
\end{equation}
providing a direct method for computing Betti numbers via kernel dimension.

A central theorem in algebraic topology states that homology groups constitute topological invariants~\cite{hatcher2005algebraic}:

\begin{theorem}[Topological invariance of homology]
If two topological spaces $X$ and $Y$ are homeomorphic, then their homology groups are isomorphic: $H_r^X \cong H_r^Y$ for all $r \geq 0$. Consequently, their Betti numbers are equal: $\beta_r^X = \beta_r^Y$.
\end{theorem}
This invariance property makes Betti numbers powerful tools for topological classification and forms the mathematical foundation for their applications in topological data analysis.

\end{document}